\documentclass[pra,superscriptaddress,amsmath]{revtex4}
\usepackage{amsthm,latexsym,amssymb,amsfonts}
\usepackage{graphicx}
\usepackage{srcltx}

\usepackage{color}

\newtheorem{theorem}{Theorem}

\newtheorem{proposition}[theorem]{Proposition}
\newtheorem{corollary}[theorem]{Corollary}
\theoremstyle{definition}
\newtheorem{lemma}[theorem]{Lemma}
\newtheorem{definition}{Definition}
\theoremstyle{remark}

\newcommand{\bra}[1]{\mbox{$\langle #1|$}}
\newcommand{\ket}[1]{\mbox{$|#1\rangle$}}
\newcommand{\ovec}[1]{\mbox{$\overrightarrow{#1}$}}

\newcommand{\op}[2]{\ket{#1}\!\bra{#2}}			
\newcommand{\ip}[2]{\left\langle #1,#2\right\rangle}
\newcommand{\tr}{\mbox{Tr}}					

\def\6{\langle}
\def\9{\rangle}
\def\tr{{\rm tr}\,}

\def\be{\begin{equation}}
\def\ee{\end{equation}}

\newcommand{\vM}{\overrightarrow{M}}

\newcommand{\Hcal}{\mathcal{H}}

\newcommand{\re}{{\mathrm{e}}}
\newcommand{\ru}{{\mathrm{u}}}

\newcommand{\rE}{{\mathrm{E}}}

\newcommand{\aD}{{\mathsf{D}}}
\newcommand{\aI}{{\mathsf{I}}}
\newcommand{\aM}{{\mathsf{M}}}
\newcommand{\aR}{{\mathsf{R}}}
\newcommand{\aT}{{\mathsf{T}}}

\newcommand{\cH}{\mathcal{H}}
\newcommand{\id}{\mathbb{I}}					

\newcommand{\rin}{{\mathrm{in}}}
\newcommand{\rout}{{\mathrm{out}}}

\begin{document}

\title{Vectorization of quantum operations and its use}

\author{Alexei Gilchrist}
\author{Daniel R. Terno} \affiliation{Department of Physics \& Astronomy, Macquarie University, Sydney NSW 2109, Australia}
\author{Christopher J. Wood} \affiliation{Institute for Quantum Computing, University of Waterloo, 200 University Av. West
Waterloo Ontario N2L 3G1, Canada}

\begin{abstract}

We give a detailed exposition of the ``vectorized" notation for dealing with quantum operations. This notation is used to highlight the relationships between  representations of completely-positive dynamics. Vectorization  considerably simplifies the analysis of different methods of quantum process tomography, and enables us to derive compact representation of the investigated quantum operations in terms of the resulting data.

\end{abstract}

\maketitle

\section{Introduction}
Quantum process tomography \cite{cn} is one of the standard tools of quantum information science, and efficient methods of processing tomographic data are of great practical and conceptual interest. Tomographic data manipulations involve massive amounts of the ``qubit/qudit algebra" on the finite-dimensional Hilbert spaces. Linear operators on them form vector spaces of their own, and the Hilbert-Schmidt inner product
\be
A\cdot B:=\tr(A^\dag B)
\ee
makes such a vector space into a Hilbert space. As a result, all the formulas that describe state evolution and measurements can be re-written in this vector form. ``Vectorized" (or ``double ket" \cite{baran}) formalism is fruitfully used in quantum information (e.g., \cite{karol,pt98,dariano00,tyson,aiello}), However, it is sometimes done on the ad hoc basis using not very transparent notation.

In this article we consistently apply it to various forms of the dynamics of open systems (Sec.~III) and quantum process tomography (Sec.~IV). It allows us to give an easy derivation of known results, and  provides a number of  new relationships between the reconstructed data and the investigated processes. To make the discussing self-contained and recognizing that a  good notation is a clue to successful derivations  we begin with a pedagogical introduction into vectorization of finite-dimensional matrices. This is a well-established area of matrix analysis \cite{hj-topics}, and part of our presentation is devoted to introducing notation which is consistent with other areas of theoretical physics and with group representation theory \cite{wkt}.

\section{Vectorization of Matrices}          \label{ssec:vecmat}
\subsection{Conventions}
We distinguish between upper (contravaraint) and lower (covariant) indices. Components of a vector $\psi\in\Hcal$ are labeled as $\psi^k$, the basis of $\Hcal$ is denoted by $\{\re_k\}$, $k=1,\ldots d\equiv \dim\Hcal$. We employ the Einstein summation convention where  the summation goes over identical upper and lower indices, hence $\psi=\sum_k\psi^k\re_k\equiv\psi^k\re_k$.

The inner product allows to identify $\re^k\equiv \re_k^{~\dag}$.  When there is a chance of confusion of the vectorial labels with the components, the former are put inside the brackets: the $a$-th component of the basis vector $k$ is denoted as
\be
(\re_k)^a=\re_{(k)}^a.
\ee
We choose the orthonormal bases, so $\re_{(k)}^a= \delta^a_k$. The Dirac notation is introduced as $|k\9\equiv \re_k$ and $\6k|\equiv \re^k$. Since
\be
\ket{\psi}=\sum_k\psi^k|k\9, \qquad \bra{\psi}=\sum_k\psi^{k*}\bra{k},
\ee
the convention for the components of general vectors is
\be
(\psi^\dag)_k=\psi^{k*}. \label{vecdual}
\ee
 Moreover, when the indices are shown explicitly, the sign of a Hermitian conjugation becomes redundant, since  positioning of the index indicates weather the object is a ket (vector) or a bra (its dual), hence $\psi_k=\psi^{k*}$.

Matrices of the size $p\times q$ represent the operators between the vector spaces of the corresponding dimensions,
\be
M:\Hcal_1\rightarrow\Hcal_2, \quad \phi^a=M^a_{~b}\psi^b,\qquad a=1, \ldots p\equiv\dim \Hcal_2, \quad b=1,\ldots,q\equiv\dim\Hcal_1.
\ee
 If the bases of the two Hilbert spaces are denoted by $\re_l$  and $\ru_k$, respectively, then
\be
M=M^k_{~l}\ru_k\re^l\equiv\sum_{k,l}M^k_{~l}|k\9\6l|\in\Hcal_2\otimes\Hcal^*_1,
\ee
where $\Hcal^*$ is the dual Hilbert space of $\Hcal$. This equation suggests a basis for the space of $p\times q$ matrices,
\be
E_{k}^{~l}\equiv\op{k}{l},
\ee
which consists of matrices with all but one  entry being zero and the unity at
 the $(kl)$-{th} entry. The matrix $M$ reads then as
\be
M=M^k_{~l}E_{k}^{~l}, \qquad k=1,\ldots p, \quad  l=1,\ldots, q.
\ee
Matrices with two lower or two upper indices can be considered as superpositions of direct product states,, as in
\be
\Psi=F^{kl}\re_k\ru_l\in\Hcal_1\otimes\Hcal_2.
\ee
In writing the matrix element $(ab)$ in any upper or lover index combination, such as $M^a_{~b}$, $M_a^{~b}$ or $(E_k^{~l})^a_{~b}$, the row index $a$ appears first.

With the above  conventions the relationships between components of a transposed matrix are given by
\be
(M^T)_{ab}=M_{ba},  \qquad (M^T)_{a}^{~b}=M^b_{~a},
\ee
the complex conjugation turns the lower into the upper indices, and vice versa,
\be
(M^*)_a^{~b}=M^{a*}_{~b},
\ee
and the Hermitian conjugation satisfies
\be
(M^\dag)^a_{~b}=M^{b*}_{~a}=(M^T)^{~b*}_{a}=(M^*)_b^{~a}
\ee
 Kronicker's delta is real and symmetric, $\delta^a_b=\delta^a_{~b}=\delta_b^{~a}$. Consistency requires to set $\delta^a_b{}^*=\delta^b_a$. We also keep in mind that $(E_a^{~b})^*=E_a^{~b}$ and $(E_a^{~b})^T=(E_a^{~b})^\dag=E_b^{~a}$, so
\be
M^*=\sum_{a,b}M^*{}_a^{~b}E_a^{~b}=M^{a*}_{~b}E_a^{~b}, \qquad M^\dag=(M^a_{~b}E_a^{~b})^\dag=\sum_{a,b}M^{a*}_{~b}E_b^{~a}=M^*{}_a^{~b}E_b^{~a}.
\ee

Consider a tensor product of two matrices, $K=M\otimes N$. A pair of row  indices $a_1$ and $a_2$  is combined into a single row index of $K$, and a pair of column indices $b_1$ and $b_2$ is combined into the new column index according to the lexicographic ordering rule that is described  below. Hence
\be
K^\alpha_{~\beta}=K^{a_1a_2}{,}_{b_1b_2}\equiv M^{a_1}_{~b_1}N^{a_2}_{~b_2},
\ee
and similarly for
\be
K^{a_1~~b_2}_{~a_2,b_1}=M^{a_1}_{~b_1}N_{a_2}^{~b_2},
\ee
etc.

\subsection{Definition and simple properties}
Representation of  matrices as vectors on a higher dimensional Hilbert space is called \emph{vectorization}. It transforms a $p\times q$ matrix $M$ into a $(pq)\times 1$ column vector denoted by $\mathrm{vec}(M)$ or $\overrightarrow{M}$ \cite{hj-topics}. We choose to do it by ordering matrix elements lexicographically, i.e. by stacking the rows of $M$ on top of each other to form a vector.   Our convention agrees with \cite{karol}, but is opposite to \cite{hj-topics}. We made this choice as to keep the same concatenation rules for both  vectorization and taking tensor products.

For example,  a $2\times 2$ matrix $M$ is re-arranged to form the four-dimensional vector $\vM$,
\be
\vM=(M^1_{~1}, M^1_{~2},M^2_{~1},M^1_{~2})^T.
\ee

To automate matrix manipulations we have to lump a pair of matrix indices $ab$, $a=1,\ldots p$, $b=1,\ldots q$ into a single vector index $\alpha$. With the stacking convention that we adopted,
\be
\alpha=f(a,b)\equiv q(a-1)+b. \label{indexall}
\ee
The inverse of vectorization restores the matrix form, as $\mathrm{mat}(\vM)=M$. The matrix indices are restored according to
\be
a=\lfloor \alpha/q\rfloor+1, \qquad b=\alpha\,\mathrm{mod}\,q\equiv \alpha-q\lfloor \alpha/q\rfloor, \label{concatind}
\ee
where $\lfloor x\rfloor$ is the largest integer less than or equal to $x$.

We  now list some useful properties of vectorized matrices~\cite{hj-topics}. For convenience we consider square matrices that act on the space $\cH$, $p=\dim\cH$. A matrix
\be
A=\sum_{kl}A^k_{~l}\op{k}{l}\equiv A^k_{~l}\re_k\re^l\in \cH\otimes\cH^*,
\ee
is transformed to the vector
\be
\ovec{A}=\sum_{kl}A^k_{~l}\re_k\otimes\re_l=\sum_l(A\re_l)\otimes\re_l
=(A\otimes\id_p)\sum_l\re_l\otimes\re_l, \label{itself}
\ee
where $\id_p$ is the identity matrix on $\cH$.

Vectorization is obviously linear: for matrices $A_\alpha$ and scalars $a^\alpha$,
        \be
        \mathrm{vec}( a^\alpha\! A_\alpha)= a^\alpha\ovec{A_\alpha}.
        \ee

Vectorization is intrinsically related to the tensor product. Our stacking convention for a matrix $M^a_{~b}$ is the same as the the index concatenation rule for elements of a tensor product space, such as $M^{ab}\mapsto M^\alpha$. This will be useful in the following. It is also easy to discover how the vectorizations of $M^T$, $M^*$, and $M^\dag$ are related to each other. Following the definition of vectorization and Eq~\eqref{vecdual}, we see that
\be
\ovec{M^*}=\ovec{M}{}^\dag, \qquad \ovec{M^\dag}=\ovec{M^T}{}^\dag. \label{vecmat}
\ee
We set $\ovec{M}$ and $\ovec{M^\dag}$ to carry the contravariant indices, so $\ovec{M^*}$ and $\ovec{M^T}$  carry the covariant ones. Hence
\be
\ovec{M}{}^\alpha\equiv M^\alpha=M^a_{~b}, \qquad \ovec{M^*}_\alpha\equiv M_\alpha={M^*}_a^{~b}=M^{a*}_{~b}=\ovec{M}{}^{\alpha*},
\ee
and
\be
 \ovec{M^\dag}{}^\alpha =M^\dag{}^a_{~b}=M^{b*}_{~a}, \qquad \ovec{M^T}_{\alpha}=M^T{}_a^{~b}=M^b_{~a}=\ovec{M^\dag}{}^{\alpha*}.
\ee

The  Hilbert-Schmidt inner product is equivalent to the usual inner product of vectors,
\be
\tr (A^\dag B)=A^*{}_b^{~a}B^b_{~a}=\ovec{A^*}_\beta \ovec{B}{}^\beta=\ovec{A}{}^\dag{}_\beta\ovec{B}{}^\beta=\ovec{A}{}^\dag\ovec{B}\equiv\6\ovec{A},\ovec{B}\9,
\ee
where $\beta=f(b,a)$.

Vectorization relates the tensor and the outer product of vectors,
\be
\ovec{A}\otimes \ovec{B}=\mbox{vec}(\ovec{A}\ovec{B^*}{}^\dag),
\ee
as follows from Eq.~\eqref{vecmat}.

To apply the summation convention we use the opposite label positioning for the basis elements $\mathrm{vec}(E_a^{~b})\equiv \ovec{\rE_\alpha}$.  Let $\alpha=f(a,b)$ and $\beta=f(k,l)$. Then the basic definition
\be
(\ovec{\rE_\alpha}){}^\beta=\delta^\beta_\alpha=\delta^k_a\delta^b_l,
\ee
and the conjugation rules for vectors and matrices lead, e.g., to
\be
(\ovec{\rE_\alpha^*})_\beta=(\ovec{\rE_\alpha}{}^\dag)_\beta=\ovec{\rE^\alpha}{}_\beta=\ovec{E_\alpha}{}^{\beta*},
\ee
which agrees with
\be
\mathrm{mat}(\ovec{\rE_\alpha^*})_k^{~l}=(E_a^{~b})^*{}_k^{~l}=(E_a^{~b})_{~l}^{k*}=(\delta_a^k\delta^b_l)^*=\delta^a_k\delta_b^l,
\ee
Similarly,
\be
\mathrm{mat}(\ovec{\rE^T_\alpha})_k^{~l}=(E_a^{~b})^l_{~k}=\delta_a^l\delta_k^b,
\ee
and
\be
\vM=M^\alpha \ovec{\rE}_\alpha, \qquad \vM^\dag=M^{\alpha*}\ovec{\rE_\alpha}{}^\dag=M_\alpha \ovec{\rE}{}^\alpha=\ovec{M^*}.
\ee

Consider  two matrices $A$ and $B$, of the sizes $p\times q$ and $q\times r$, respectively. Their product $C=AB$ can be written as a vector $\ovec{AB}$ in two ways. The  matrix $C$  results from the left action of $A$ on $B$,
so in a vectorized notation
\be
C^\alpha={\aM_A}^\alpha_{~\beta} B^\beta,
\ee
where the matrix $\aM_A$ will be now determined. Writing the indices in full and using matrix multiplication rules leads to
\be
C^a_{~b}={\aM_A}^{a~~d}_{~b,c}B^c_{~d}=A^a_{~c}B^c_{~d}\id^d_{~b}=A^a_{~c}\id^{~d}_{b}B^c_{~d}, \label{transa}
\ee
so
\be
{\aM_A}^{a~~d}_{~b,c}=A^a_{~c}\delta^{~d}_b.
\ee
As a result,
\be
\ovec{AB}=(A\otimes\id_r)\ovec{B}. \label{lefta}
\ee
Now we see that Eq.~\eqref{itself} for $\ovec{A}$ is  a corollary of the above result with  $B\mapsto\id$. The matrix $C$ also results from the right action of $B$ on $A$, which is expressed as
\be
\ovec{AB}=(\id_p\otimes B^T)\ovec{A}. \label{righta}
\ee

The following lemma \cite{hj-topics} deals with the triple product of matrices:
\begin{lemma}
Let $A, B, C$ and $X$ be   $p\times q$,  $r\times s$, $p\times s$  and $q\times r$ matrices, respectively. Then the matrix equation
\be
AXB=C
\ee
for $qs$ unknowns $X^l_{~m}$ is equivalent to the system of $ps$ equations
\be
(A\otimes B^T)\ovec{X}=\ovec{C}.
\ee
That is to say,
\be
\ovec{AXB}=(A\otimes B^T)\ovec{X}.
\ee
The proof consists in the repeated application of Eqs.~(\ref{lefta}) and (\ref{righta}),
\be
\ovec{AXB}=(A\otimes\id_s)\ovec{XB}=(A\otimes\id_s)(\id_q\otimes B^T)\ovec{X}=(A\otimes B^T)\ovec{X}.
\ee
\hfill $\square$
\end{lemma}


\subsection{Reshuffling}\label{ssec:reshuffling}

The  SWAP operation  changes the order of subsystems in a tensor product:
\be
|\psi\9_1\otimes|\phi\9_2\mapsto \textrm{SWAP}(|\psi\9_1\otimes|\phi\9_2)=|\phi\9_2\otimes|\psi\9_1.
\ee
For two identical systems this operation swaps their quantum states.  Consider two vectors $\psi\in\cH_1$ and $\phi\in\cH_2$. In the component form we write them as
\be
\psi=\psi^{a_1}\re_{a_1}, \qquad a_1=1,\ldots, p=\dim\cH_1,
\ee
 and
\be
\phi=\psi^{a_2}\re_{a_2}, \qquad a_2=1,\ldots, r=\dim\cH_2,
\ee

 A single index $\alpha$ of the tensor product $\Psi^\alpha=(\psi\otimes \phi)^\alpha$ is built from the indices of its subsystems as
in Eq.~\eqref{concatind},
\be
\alpha=r(a_1-1)+a_2.
\ee
Similarly,  a single vectorial index for $\Phi=\phi\otimes\psi$ is
\be
\beta=q(a_2-1)m+a_1.
\ee
The bases of the two tensor products are certain permutations of each other. It is easy to see that this permutation $\beta=\sigma[r,p](\alpha)$ is given by the table
\be
\left(\begin{array}{cccccc|ccc|ccc}
      1 & 2  &\ldots &   s     &\ldots & r        &  r+1    & r+2  &\ldots &(p-1)r+1 & \ldots & pr\\
      1 & 1+p&\ldots &1+(s-1)p &\ldots & 1+(r-1)p &  2     & 2+p  &\ldots &p        &\ldots  & p+(r-1) p  \end{array}\right), \label{permut}
\ee
where the first row contains  indices of the components of $\Psi$ that are matched with the corresponding indices of $\Phi$ in the second row. The vertical lines separate the first and the last $r$ components of $\Psi$ from the rest, and  $s=1,\ldots r$. Its inverse, $\sigma[r,p]^{-1}$ associates to each element of $\Phi$ an element of $\Psi$ and is obtained by interchanging the rows. As a result, if we label the SWAP operator as $S(r,p)$, then
\be
(\phi\otimes\psi)^\alpha=S(r,p)^\alpha_{~\beta}(\psi\otimes \phi)^\beta
\ee
where
\be
S(r,p)^\alpha_{~\beta}=\delta^{\sigma[r,p](\alpha)}_{~\beta}.
\ee
It follows from this construction that $S$ is an orthogonal matrix, and $S(r,p)^T=S(r,p)^{-1}=S(p,r)$. For example, a qubit SWAP is
\be
S(2,2)= \left(
\begin{array}{cccc}
1 & 0 & 0 & 0 \\
0 & 0 & 1 & 0  \\
0 & 1 & 0 & 0  \\
0 & 0 & 0 & 1 \\
\end{array}\right),
\ee
and the SWAP of $2\times 3$ systems is done with
\be
S(2,3)= \left(
\begin{array}{cccccc}
1 & 0 & 0 & 0 & 0 & 0 \\
0 & 0 & 0 & 1 & 0 & 0 \\
0 & 1 & 0 & 0 & 0 & 0 \\
0 & 0 & 0 & 0 & 1 & 0 \\
0 & 0 & 1 & 0 & 0 & 0 \\
0 & 0 & 0 & 0 & 0 & 1
\end{array}\right).
\ee
If we restore the original indices we see that SWAP can be treated as a matrix transposition
\be
\Phi^{a_2a_1}=\Psi^{a_1a_2}=S^{a_2a_1}_{~~~~,b_1b_2}\Psi^{b_1b_2},
\ee
and
\be
S^{a_2a_1}_{~~~~,b_1b_2}=\delta^{a_2}_{b_2}\delta^{a_1}_{b_1}.
\ee
For two square matrices $M$ and $N$ that act on the spaces $\cH_1$ and $\cH_2$, respectively,  the SWAP operation results in
\be
N\otimes M = S(r,p) (M\otimes N) S(r,p)^T.
\ee

As we will see in the next sections, it is  important to relate $\mathrm{vec}(M)\otimes\mathrm{vec}(N)$ and $\mathrm{vec}(M\otimes N)$. It is done by the operation of \textit{reshuffling} \cite{karol}. Let the matrices $M$ and $N$ be of the sizes $p\times q$ and $r\times s$, respectively. A matrix element
\be
M^{a_1}_{~b_1}N^{a_2}_{~b_2}
\ee
can be interpreted either according to `` first vectorize, then tensor", as
\be
M^{a_1}_{~b_1}N^{a_2}_{~b_2}=:C^{a_1~a_2}_{~b_1,~b_2}\stackrel{\mathrm{vec}}{\mapsto} C^{\alpha\beta}\stackrel{\otimes}{\mapsto} C^A=\left(\mathrm{vec}(M)\otimes\mathrm{vec}(N)\right)^A,
\ee
or as
\be
M^{a_1}_{~b_1}N^{a_2}_{~b_2}=:\overline{C}{}^{a_1 a_2}_{~~~~,b_1 b_2}\stackrel{\otimes}{\mapsto} \overline{C}{}^\gamma_{~\delta}\stackrel{\mathrm{vec}}{\mapsto} \overline{C}{}^B=\mathrm{vec}(M\otimes N)^B,
\ee
according to ``first tensor, then vectorize" precept.

A repeated application of the index concatenation definition gives
\be
A=rsq(a_1-1)+rs(b_1-1)+s(a_2-1)+b_2, \qquad B=rsq(a_1-1)+qs(a_2-1)+s(b_1-1)+b_2.
\ee
Hence swapping the indices $a_2$ and $b_1$ brings $\ovec{M}\otimes\ovec{N}$ to $\ovec{M\otimes N}$.  This can be formalized by the following

\begin{definition}[Reshuffling Matrix]
For matrices $M$ and $N$ of the size $p\times q$ and $r\times s$, respectively,  the reshuffling matrix $\aR(M,N)$ is defined by
\be
\aR(M,N)\equiv\aR(p,q,r,s)=\id_p\otimes S(r,q)\otimes\id_s,
\ee
\end{definition}
\noindent Our discussion established the
\begin{proposition}[Reshuffling]
Let the matrices $M$ and $N$ be of the size $p\times q$ and $r\times s$, respectively. Then
\be
\mathrm{vec}(M\otimes N) = \aR(p,q,r,s)(\ovec{M}\otimes\ovec{N}). \label{vecten}
\ee
\end{proposition}

\bigskip

 A Mathematica package by Miszczak,  Gawron, and Pucha{\l}a \cite{pack} for the analysis of quantum states and operations implements matrix vectorization and resufling for the use in the relevant functions \footnote{Note that in \cite{pack} our vectorization operation is called ``reshaping", and the name ``vectorization" refers to the column-by-column stacking, as in \cite{hj-topics}.}.
\section{Applications of Vectorization to Open Quantum Systems}

The simplest use of the vectorization is in analyzing the convex probability domains of quantum measurements \cite{pt98}.  Consider a positive operator valued measure (POVM) with $N$ outcomes. Vectorization makes it nearly obvious that the domain in the probability space formed by probabilities of all outcomes with all possible states of a $n$ dimensional system has at most $n^2-1$ dimensions.

\subsection{Completely Positive Maps}     \label{ssec:cpaxioms}
Evolution of an open quantum system whose initial state is uncorrelated with the environment is described by a completely positive map. Label the initial state of the system ($\cH_1$) as $\rho$, and the state of  its environment ($\cH_2$) by $\omega$. If  the joint state is given by $\tau_{12}=\rho\otimes\omega$, and the overall unitary evolution is $U_{12}$, then the final state of the system is given by
\be
\rho \mapsto \aT(\rho)=\tr_{\cH_2}(U_{12}\rho\otimes\omega U_{12}^\dag), \label{bighilb}
\ee
where the operation of partial tracing is defined by
\be
(\tr_2\tau)^{a_1}_{~b_1}=\tau^{a_1 a_2}_{~~~~,b_1a_2}.
\ee
 Such a map $\aT$ is trace-preserving, convex-linear and completely positive \cite{karol}. This latter property means  that $\aT$ is positive (maps positive matrices to positive matrices), and also if we introduce an auxiliary system of arbitrary dimension then the map $\aT\otimes\aI$ on the joint system is positive, where $\aI$ is the identity map on the auxiliary system.

%
%

A map satisfying these three axioms is referred to as a completely positive, trace preserving (CPTP) map. It is possible to relax the trace condition to $\tr[\aT(\rho)]\le\tr[\rho]$, allowing for completely positive trace decreasing maps. We will only be concerned with CPTP maps and we may refer to maps satisfying all three of these axioms simply as completely-positive maps (CP maps).


Any CPTP map has a convenient operator sum representation:

\begin{theorem}[Kraus representation]\label{thm:kraus}
A map $\aT$ acting on density operators of $\Hcal$ is CPTP if and only if there exists a set of bounded operators $\{K_n\}$ acting on $\Hcal$ such that
\begin{equation}
\aT(\rho) = \sum_n K_n \rho K_n^\dagger
\quad \mbox{ where }\quad \sum_n K_n^\dagger K_n = \id.
\label{eqn:krausform}
\end{equation}
\end{theorem}
\noindent  The operators $K_n$ are called \emph{Kraus matrices} and they satisfy $\sum_n K_n^\dagger K_n = \id$, which is known as the  \emph{completeness relation}.
A Kraus representation of a given process $\aT$ is not unique. This can be useful as different system-environment interactions may still give rise to the same reduced dynamics on the system.

In general,
a linear hermiticity-preserving transformation $\aT$ (completely positive or not) acting on the
space  of density matrices may be represented by the
{\emph{dynamical (Choi) matrix}} $\aD(\aT)$ \cite{karol, su61,choi},
\begin{equation}
 \rho'{}^a_{~b}=\aD^{a~~~d}_{~c,b}\rho^c_{~d},\label{introchoi}
\end{equation}
and a vectorized version of this relation
\be
\ovec{\rho}'^\alpha=\bar{\aD}^\alpha_{~\beta}\ovec{\rho}^\beta, \label{dynres}
\ee
uses the reshuffled matrix $\bar{D}$,
\be
\bar{\aD}^{a~~~d}_{~b,c}=\aD^{a~~~d}_{~c,b},
\ee
 $\mathrm{vec}(\bar{\aD})=\aR \ovec{\aD}$.

The dynamical matrix has a number of useful properties \cite{karol}. The trace
preserving condition  is equivalent to the constraint  on the partial trace of the dynamical matrix,
\begin{equation}
\aD^{a~~~d}_{~c,a}= \delta^d_c \ ,
\end{equation}
which implies that the eigenvalues sum up to the system's dimension, $\sum_a\lambda_a=d_1$.   Moreover, if the map is unital,
i.e., it maps the maximally mixed state into the
 maximally mixed state, then
\begin{equation}
\aD^{a~~~s}_{~s,b}=\delta^a_b \ .
\end{equation}

The dynamical matrix is Hermitian, $\aD^\dag{}^\alpha_{~\beta}=\aD^\alpha_{~\beta}$, and
due to a theorem of Choi \cite{choi}
its positivity is equivalent to the complete positivity of $\aT$:
\begin{theorem}[Dynamical Matrix]\label{thm:process}
A quantum operation $\aT$ on a $d$-dimensional system $S$ is CP if and only if its dynamical matrix $\aD(\aT)$ is positive-semidefinite $(\aD(\aT)\ge0)$.
\end{theorem}
The proof is based on the eigendecomposition of the Choi matrix,
\begin{equation}
\aD^{a~~~d}_{~c,b}=\aD^\alpha_{~\beta}=\sum_n\lambda_n \ovec{M_n}{}^\alpha\ovec{M_n}{}^\dag_\beta=\sum_n\lambda_n\ovec{M_n}{}^\alpha\ovec{M^*_n}_\beta=\sum_n\lambda_n M_n{}^a_{~c} M_n^*{}^{~d}_{b}, \label{eigenD}
\end{equation}
where $\alpha=f(a,c)$, $\beta=f(b,d)$, and if all the eigenvalues $\lambda_n$ are non--negative it is possible to define Kraus matrices  by absorbing the eigenvalues,
\be
K_n=\sqrt{\lambda_n}M_n.
\ee
We also get a compact expression for the reshuffled matrix,
\be
\bar{\aD}=\sum_n K_n\otimes K^*_n. \label{bartensor}
\ee

\subsection{Linear Superoperator}     \label{ssec:superop}

Vectorization allows a different perspective on Choi matrix and its reshuffled version. We consider a combination of basis matrices $E_a^{~b}$ and $\aT(E_a^{~b})\equiv E'{}_a^{~b}$, with  suitably arranged component indices.

A \emph{linear superoperator} $\Phi_\aT$ acts on vectorized density matrices and is defined by
\begin{equation}
\Phi_\aT\ovec{\rho}\equiv\ovec{\aT(\rho)}.
\label{eqn:sopevo}
\end{equation}
On the one hand, since
\be
\6\rE_\alpha,\rE_\beta\9=\ovec{\rE^\alpha}\,\ovec{\rE_\beta}=\tr(E_a^{~b}{}^\dag E_c^{~d})=\delta^a_c\delta_b^d=\delta^\alpha_\beta, \label{phi1}
\ee
 we have
\be
\rho^\beta\ovec{\rE'_\alpha}\ovec{\rE^\alpha}\ovec{\rE_\beta}=\rho^\beta\ovec{\rE'_\beta}=\Phi_\aT\ovec{\rho}. \label{linact}
\ee
As a result,
\begin{equation}
\Phi_\aT=\sum_\alpha\ovec{E'_\alpha}\,\ovec{\rE_\alpha}{}^\dag=\ovec{E'_\alpha}\,\ovec{\rE^\alpha}.
\label{eqn:sodef}
\end{equation}
A comparison with Eq.~\eqref{dynres} identifies the linear superoperator with the reshuffled dynamical matrix, $\Phi_\aT=\bar{\aD}(\aT)$. The matrix elements of $\Phi_\aT$ are obtained by
\be
\Phi_\aT{}^\kappa_{~\mu}=\ovec{E'_\alpha}{}^\kappa\ovec{\rE^\alpha}_\lambda=\sum_{a,b}(E'{}_a^{~b})^k_{~l}(E{}_a^{~b}{}^*)_m^{~n}
=(E'{}_a^{~b})^k_{~l}\delta_m^a\delta_b^n=(E'{}_m^{~n})^k_{~l}=(E'_\mu)^\kappa.
\ee
Indeed, $\rho'_\kappa=\Phi_\aT{}^\kappa_{~\mu}\rho^\mu$, so
\be
\bar{\aD}^{k~~~n}_{~l,m}=(E'{}_m^{~n})^k_{~l}=\sum_x (K_x)^k_{~m} (K^*_x)_{l}^{~n}.
\ee
We note that when expressed in terms of the Kraus matrices $K_n$  the reshuffled matrix $\bar{\aD}$ is built according to ``first tensor, then vectorize" prescription. However, the opposite approach ``first vectorize, then tensor"  is taken when $\ovec{\Phi_\aT}$ is expressed in terms of the basis matrices and their transforms.

\subsection{Process matrix}

We define the  \textit{process matrix} $\Lambda_\aT$ for a  map $\aT$ as a matrix on a fictitious space $\cH_1\otimes\cH_2$, where the two spaces are copies of $\cH$. Without going through the vectorization of Eqs.~\eqref{phi1}, \eqref{linact}, the transformation $\rho'=\aT(\rho)$ can be written as
\be
\rho'=\sum_{a,b}\rho^c_{~d}E'{}_a^{~b}\,\tr(E_a^{~b}{}^\dag E_c^{~d})=\rho^c_{~d}E'{}_a^{~b}\,\tr(E_b^{~a} E_c^{~d}).
\ee
Thus it is given by
\begin{equation}
\aT(\rho)=\tr_2\left[\Lambda_\aT^{T_2}(\id\otimes\rho)\right], \qquad \Lambda_\aT^{T_2}\equiv E'{}_a^{~b} \otimes E_{b}^{~a}. \label{eqn:processmap}
\end{equation}
For our purposes it is more convenient to use  the matrix arrangement of $\Phi_\aT$, and define a new four-indexed object according to the ``first tensor, then vectorize" prescription,
\be
\Lambda_\aT\equiv E'{}_a^{~b} \otimes E_{b}^{~a}{}^\dag=\sum_{a,b} E'{}_a^{~b} \otimes E_{a}^{~b}{}^*,
\ee
where we keep a complex conjugation as a reminder for the correct index placement. Then
\be
\aT(\rho)=\tr_2\left[\Lambda_\aT(\id\otimes\rho^T)\right],
\ee
and
\be
\Lambda_\aT{}^{k~~~n}_{~l,m}=(\sum_{a,b} E'{}_a^{~b} \otimes E_{a}^{~b}{}^*){}^{k~~~n}_{~l,m}=(E'{}_a^{~b})^k_{~m}\delta_l^a\delta_b^n=(E'{}_l^{~n})^k_{~m}.
\ee
A comparison with Eq~\eqref{introchoi} shows that $\aD(\aT)=\Lambda_\aT$.

\subsection{Summary of relationships}

The Jamio{\l}kowski isomorphism \cite{karol, jamio} identifies dynamical matrices of CP maps with certain entangled states. Consider a generalization of the Bell state $|\Phi^+\9=(|++\9+|--\9)/2$ on the space $\cH_1\otimes\cH_2$ of the previous section. Its density matrix is given by
\be
\tau_+=\sum_{a,b} E_{a}^{~b}\otimes E_{a}^{~b}/d,
\ee
where $d=\dim\cH$. It is now easy to write the matrix elements of $\rho'=(\aT\otimes\aI)\rho$,
\be\tau'_+{}^{kl}_{~~,mn}=(E'{}_l^{~n})^k_{~m}/d=\Lambda_\aT{}^{k~~~n}_{~l,m}/d,
\ee
which establishes
\begin{theorem}[Jamio{\l}kowski isomorphism]
Any linear map $\aT$ acting on the space of mixed states on the Hilbert space $\cH$
 can be associated, via its dynamical matrix $\aD(\aT)$, with an operator on the enlarged Hilbert spacee $\cH\otimes\cH$,
\be
\aD(\aT)\cong(\aT\otimes\aI)\tau_+=\tau'_+,
\ee
where for a completely positive map $\aT$ the matrix $\tau'_+$ is a valid quantum state.
 \end{theorem}
As a  computational aid we present a compact summary of the relationships between different representations of a completely positive map $\aT$.

\begin{table}[htbp]
\begin{center}
\begin{tabular}{|c||c|c|c|}
  \hline
To$\backslash$From & $\aD(\aT) $	& $\Phi_\aT$ 						& $ K_n $\\
\hline\hline
$\aT (\rho)=$		&  $\tr_2\left[\aD(\id_1\otimes\rho^T)\right] $& $\mathrm{mat}(\Phi_\aT\ovec{\rho})$ &  $\sum_n K_n\rho K_n^\dagger$\\
\hline
$\aD =$ 	& $\aD$ 		& $\mathrm{mat}(\aR^{-1}\ovec{\Phi_\aT})$ 	&  $\sum_{n} \ovec{K_n}\ovec{K_n}{}^\dag$  \\
\hline
$\Phi_\aT =$    	& $\mathrm{mat}(\aR\ovec{\aD})$   & $\Phi_\aT$ 	&   $\sum_n K_n\otimes K_n^*$ \\
\hline
$K_n =$           & $\sqrt{\lambda_n}\mathrm{mat}(\ovec{M_n})$ & $\longleftarrow$ & $K_n$ \\
  \hline
\end{tabular}
\end{center}
\caption{Relationships between mathematical representations of CP maps.}
\label{tab:cpreps}
\end{table}
Two possible viewpoints on both $\aD$ and $\Phi_\aT$ as obtained either according to ``first tensor, then vectorize" or ``first vectorize, then tensor" approaches (with a counterpart being obtained by the other procedure) are consistent, because the reshuffling matrix involved in the process satisfies $\aR=\aR^{-1}$. The latter result follows from Eq.~\eqref{permut}, with all the dimensions being equal to $\dim\cH$.

\section{Process Tomography using the Superoperator representation}

\subsection{Dual bases}

Quantum process tomography is used to characterize the properties of quantum  information-processing devices. The standard process tomography \cite{cn} uses a tomographically complete set of input states that spans the entire set of states. The input states are sent through the investigated device and its action is then reconstructed by analyzing the outputs.  A  set of dual states $D=\{ D_\nu \}$  for a tomographically complete set  $\rho_{\rin}=\{ \rho_\mu\}$, $\mu=1,\ldots, n=d^2$, where $d=\dim\cH$, plays an important role in the mathematics of reconstruction. It is defined by the orthogonality relation $\tr (D_\nu^\dagger \rho_\mu ) =\delta_{\nu\mu}$, $\forall \mu,\nu$.

If we rewrite this in a vectorized notation we obtain
\be
\ip{\ovec{D_\mu}}{\ovec{\rho_\nu}}= \delta_{\mu\nu}
\ee
hence
\be
\ovec{\rho_\mu}\,\ovec{\rho^\mu}\ovec{D_\nu}=\ovec{\rho_\mu}\delta^\mu_\nu=\ovec{\rho_\nu}
\ee
Introducing the operator
\be
P=\sum_\mu \ovec{\rho_\mu}\,\ovec{\rho_\mu}{}^\dag\equiv\ovec{\rho_\mu}\,\ovec{\rho^\mu},
\ee
we see that $P\ovec{D_\nu}=\ovec{\rho_\nu}$, hence
\be
\ovec{D_\nu}=P^{-1}\ovec{\rho_\nu}.
\ee

It is possible to write this relationship between a tomographic set and its dual in a simpler and computationally more efficient way.  We write vectors of the set $\rho_\rin$  as the columns of a  matrix
\be
[\rho_\rin]=(\ovec{\rho_1},\ldots,\ovec{\rho_n}). \label{eqn:matvec}
\ee
Since its columns are linearly independent, this $d^2\times d^2$ matrix is invertible. Its Hermitian conjugate can be written as
\begin{eqnarray}
[\rho_{\rin}]^\dagger &=&\left(
\begin{array}{c}
\ovec{\rho_1}{}^\dag \\
\vdots \\
\ovec{\rho_n}{}^\dag
\end{array}
\right).
\end{eqnarray}
With this notation we get
\be
P=[\rho_\rin][\rho_\rin]^\dag.
\ee
As a result,
\be
\ovec{D_\mu} =[\rho_{\rin}]^{-1}{}^\dagger[\rho_{\rin}]^{-1}\ovec{\rho_\mu}.
\ee
If we introduce $[D]=(\ovec{D_1},\ldots,\ovec{D_n})$, then an even simpler expression is obtained:
\be
[D]=[\rho_{\rin}]^{-1}{}^\dagger[\rho_{\rin}]^{-1}[\rho_{\rin}]=[\rho_{\rin}]^{-1}{}^\dag.
\ee

Now we apply this to the standard process tomography.

\subsection{Process tomography with $\Phi_\aT$}

\begin{proposition}
Let $\rho_{\rin}=\{\rho_\mu\}$ be a tomographically complete set of input states with dual basis $D=\{D_\mu\}$. Then the linear superoperator $\Phi_\aT$ for a CP map $\aT: \rho \mapsto \aT(\rho) $ is given by
\be
\Phi_\aT = \sum_\mu \ovec{\aT(\rho_\mu)}\,\ovec{D_\mu}{}^\dag\equiv \ovec{\rho'_\mu}\,\ovec{D^\mu} \label{phinew}
\ee
\begin{proof}
Consider an arbitrary state $\rho= p^\mu\rho_\mu$. We have to show $\Phi_\aT \ovec{\rho}=\mathrm{vec}{\,\aT(\rho)}$. Applying the above expression to $\ovec{\rho}$ we see that
\be
\Phi_\aT(\ovec{\rho})=p^\nu \ovec{\rho'_\mu}\ip{D_\mu}{\rho_\nu}=p^\nu \ovec{\rho'_\mu}\delta_\nu^\mu=p^\mu\ovec{\rho'_\mu}=\aT(\rho).
\ee
\end{proof}
\end{proposition}

We simplify this expression by using the matrix of vectorized input states, Eq.~\eqref{eqn:matvec}. This leads us to the following
\begin{proposition}
For the set of output states $\rho_{\mathrm{out}}=\{\aT(\rho_\mu)\}$, where $\rho_{\rin}=\{\rho_\mu\}$ is a tomographically complete set of input states, the linear superoperator  $\Phi_\aT$ is given by
\be
\Phi_\aT = [\rho_{\rout}][ \rho_{\rin}]^{-1}
\ee
\begin{proof} We use Eq.~\eqref{phinew}.
\be
[\rho_{\rout}][ \rho_{\rin}]^{-1}=[\rho_{\rout}][D]^\dag=\ovec{\rho'_\mu}\,\ovec{D^\mu}=\Phi_\aT.
\ee
\end{proof}
\end{proposition}

We rewrite this result in terms of the  probabilities of various experimental outcomes.

\begin{proposition}
Let $\{M_\mu\}$ be a tomographically complete measurement set ($M_\mu\geq 0$, $\sum M_\mu=\id$) with a dual basis $\{E^\nu\}$, $\tr (E_\nu^\dag M_\mu)=\delta^\nu_\mu$.  Then
\be
\Phi_\aT = \left[ E \right]\left[ m \right]\left[ D \right]^\dagger, \label{spt}
\ee
where $[E]=(\ovec{E_1},\ldots, \ovec{E_n})$ is a matrix of the vectorized dual elements, and $[m]=(m_{\mu\nu})$ is a matrix of probabilities, $m_{\mu\nu}=\tr (M_\mu^\dag\rho'_\nu)$.
\begin{proof}
Since  any $\rho$ is reconstructed according to $\rho=E^\mu m_\mu$, where $m_\mu=\tr(M_\mu\rho)$,
\be
[\rho_{\rout}]= [E][m],
\ee
and $[D]^\dag=[\rho_\rin]^{-1}.$
\end{proof}
\end{proposition}

\subsection{Ancilla-assisted process tomography with linear superoperator}

Presentation of Jamio{\l}kowski isomorphism and the manipulation of data in the ancilla-assisted quantum tomography (AAPT) \cite{dariano00,alte} also benifit from the vectorized notation.

We introduce an auxiliary system (ancilla) $\cH_2$ to our principal system $\cH_1$, so that the state space of the joint system is given  by $\Hcal_1\otimes\Hcal_2$.  AAPT aims to reconstruct the CP map $\aT$ on the states of $\cH$  the action of $\aT\otimes\aI$ on a single state $\tau_{12}$ of this combined system.
Evolution on $\cH_1$ results in an  operation $\aT$ described by a superoperator $\Phi_\aT$, and the ancilla does not evolve.
Any initial state $\tau_{12}$ of the joint system can be represented as
\be
\tau_{12}=\sum_\mu w^\mu\rho_\mu\otimes\omega_\mu, \qquad \sum_\mu w^\mu=1
\ee
which is in general entangled (it is separable if and only if all $w^\mu\geq 0$).

\begin{lemma}
The joint system dynamical matrix has the form
\be
\Phi_{\aT\otimes\aI}=\aR(\Phi_\aT\otimes\bar{\aI})\aR^{-1},
\ee
where $\aR$ is the reshuffling matrix, which in this case again satisfies $\aR=\aR^{-1}$, and the linear superoperator of the identity map is $\bar{\aI}\equiv\Phi_\aI$.
\begin{proof}
Linearity of the evolution is expressed as
\be
\tau'=(\aT\otimes\aI)\tau=\sum_\mu w^\mu \aT(\rho_\mu)\otimes\omega_\mu.
\ee
Hence
\be
\ovec{\tau}'=\sum_\mu w^\mu\ovec{\rho'_\mu\otimes\omega_\mu}=\sum_\mu w^\mu\aR(\ovec{\rho_\mu}'\otimes\ovec{\omega_\mu})
=\aR(\Phi_\aT\otimes\bar{\aI})\sum_\mu w^\mu \ovec{\rho_\mu}\otimes\ovec{\omega_\mu}.
\ee
Matrix elements  of $\bar{\aI}$ satisfy $\bar{\aI}^{k~~~n}_{~l,m}=\delta^k_m\delta^n_l=\id^{\alpha}_{~\beta}$. Another reshuffling leads to the desired result,
\be
\ovec{\tau}'=\aR(\Phi_\aT\otimes\bar{\aI})\aR^{-1}\ovec{\tau}.
\ee
\end{proof}
\end{lemma}
Both vectors
\be
\ovec{\tau_\rin}^A=\ovec{\tau_{\rin}}{}^{\alpha\beta}\equiv(\aR^{-1}\ovec{\tau})^{\alpha\beta}=(\aR^{-1}\ovec{\tau})^{k~m}_{~l,~n}=
\sum_\mu w^\mu\rho_\mu{}^k_{~l}\,\omega_\mu{}^m_{~n},
\ee
and $\ovec{\tau_\rout}{}^A\equiv(\aR^{-1}\ovec{\tau}')^{\alpha\beta}$ correspond to the right hand side of Eq.~\eqref{vecten}. Their relationship is of the forms of Eq.~\eqref{lefta}:
\be
\ovec{\tau_\rout}{}^{\alpha\beta}=(\Phi_\aT{}^\alpha_{~\gamma}\id^{\beta}_{~\delta})\ovec{\tau_\rin}^{\gamma\delta}.
\ee
Introducing $\Phi_\tau\equiv\mathrm{mat}(\ovec{\tau_\rin})$ we have
\be
\ovec{\tau_\rout}=(\Phi_\aT\otimes\id)\ovec{\tau_\rin}=(\id\otimes\Phi_\tau^T)\ovec{\Phi_\aT}.
\ee
Hence we can recover the dynamical matrix $\Phi_\aT$ from the output state $\tau_{AS}$ when the matrix $\Phi_\tau$ is invertible:

\begin{proposition}
A linear superoperator $\Phi_\aT$ is recovered from the output of AAPT with the initial state $\tau$ according to
\be
\ovec{\Phi_\aT}{}^\alpha_{~\beta}=(\id\otimes(\Phi_\tau^{-1})^T)^\alpha_{~\beta,\gamma\delta}\ovec{\tau_\rout}^{\gamma\delta} \label{aapt}
\ee
\end{proposition}

An important special case is the entanglement-assisted process tomography, where  the input state is a maximally entangled
\be
\tau=\tau_+=\sum_{ij}\op{i}{j}\otimes\op{i}{j}/d.
\ee
It corresponds to $\Phi_{\tau_+}=\mathrm{mat}(\ovec{\tau_+})=\id$. Hence we established a useful expression for the dynamical matrix and a dual form of the Jamio{\l}kowski isomorphism
\begin{corollary}[entanglement-assisted process tomography]
 In the entanglement assisted process tomography (AAPT with the maximally entangled initial state $\tau_+$) the dynamical matrix the linear superoperator $\Psi_\aT$ is determined by the output state according to
 \be
\ovec{\Phi_\aT}=\ovec{\tau_\rout}. \label{eapt}
 \ee
\end{corollary}

\section{Conclusions and outlook}
Adopting vectorized notation allows transparent and consistent representation of various forms of open system dynamics, isomorphism between states and operations and representation of process tomography. Neat expressions for process tomography \eqref{spt}, \eqref{aapt} and \eqref{eapt} use already reconstructed output states. Processing of actual state tomographic data  is much more involved. In particular,  relative frequencies cannot be directly taken  as probabilities \cite{pt98, alte}, assumption of completely positive dynamics should be justified or may not be true \cite{lidar,us}, matrices may have only generalized inverses \cite{dariano09}. A transparent and versatile notation is a great asset in dealing with these issues, and we expect that it will simplify some of the existing formulas and bring to light new useful relationship, similarly to presented in this work.

\acknowledgments
We thank Karol \.{Z}yczkowski for many useful discussions. A correspondence with  Andrea Aiello, Joseph Emerson, Jaroslaw Miszczak,  Yutaka Shikano, and Jon Tyson is gratefully acknowledged.  The work of DRT was supported in part by the grant from  the Australian Academy of science.  CJW was supported by the Perimeter Scholars International program.

\end{document}